\documentstyle[aas2pp4,psbox]{article}
\slugcomment{To appear in ApJ}
\lefthead{J. Yokogawa et al.}
\righthead{New Transient X-ray Pulsar in the SMC}

\begin{document}

\title{{\it ASCA} Observation of the New Transient X-ray Pulsar
XTE~J0111.2--7317 in the Small Magellanic Cloud}

\author{Jun~Yokogawa\altaffilmark{1}, 
Biswajit~Paul\altaffilmark{2,}\altaffilmark{3}, 
Masanobu~Ozaki\altaffilmark{2},
Fumiaki~Nagase\altaffilmark{2}, 
Deepto~Chakrabarty\altaffilmark{4}, and
Toshiaki~Takeshima\altaffilmark{5,}\altaffilmark{6}}

\altaffiltext{1}{Department of Physics, Graduate School of Science,
Kyoto University, Sakyo-ku, Kyoto 606-8502, Japan; 
jun@cr.scphys.kyoto-u.ac.jp}

\altaffiltext{2}{Institute of Space and Astronautical Science, 3-1-1 Yoshinodai, 
Sagamihara, Kanagawa 229-8510, Japan; 
bpaul@astro.isas.ac.jp, ozaki@astro.isas.ac.jp, nagase@astro.isas.ac.jp}

\altaffiltext{3}{On leave from the Tata Institute of Fundamental Research, 
Homi Bhabha Road, Mumbai, 400005, India}

\altaffiltext{4}{Department of Physics and Center for Space Research, 
Massachusetts Institute of Technology, Cambridge, MA 02139, USA; 
deepto@space.mit.edu}

\altaffiltext{5}{Laboratory for High Energy Astrophysics, NASA,
Goddard Space Flight Center, Greenbelt, MD 20771, USA; 
takeshim@ginpo.gsfc.nasa.gov}

\altaffiltext{6}{Universities Space Research Association, Seabrook, 
MD 20706, USA}

\begin{abstract}
The new transient X-ray pulsar XTE~J0111.2--7317 was observed with 
{\it Advanced Satellite for Cosmology and Astrophysics} ({\it ASCA})
on 1998 November 18, a few days after its discovery with the Proportional
Counter Array onboard the {\it Rossi X-ray Timing Explorer}. 
The source was detected at a flux level of 
$3.6\times10^{-10}$~erg~cm$^{-2}$~s$^{-1}$ in the 0.7--10.0~keV band, 
which corresponds to the X-ray luminosity of 
$1.8\times10^{38}$~erg~s$^{-1}$, 
if a distance of 65 kpc for this pulsar in the Small Magellanic Cloud is assumed.  
Nearly sinusoidal pulsations with a
period of $30.9497 \pm 0.0004$~s were unambiguously detected during the {\it
ASCA} observation. The pulsed fraction is low and slightly energy dependent with average
value of $\sim27$\%. 
The energy spectrum shows a large soft excess below $\sim 2$~keV 
when fitted to a simple power-law type model. 
The soft excess is eliminated if the spectrum is fitted to 
an ``inversely broken power-law'' model, 
in which photon indices below and above a break energy of 1.5~keV
are 2.3 and 0.8, respectively.
The soft excess can also be described by a blackbody or a thermal bremsstrahlung 
when the spectrum above $\sim 2$~keV is modeled by a power-law. 
In these models, however, 
the thermal soft component requires a very large emission zone, 
and hence it is difficult to explain the observed 
pulsations at energies below 2~keV. 
A bright state of the source enables
us to identify a weak iron line feature at 6.4~keV with an equivalent width of
$50\pm14$~eV. 
Pulse phase resolved spectroscopy revealed 
a slight hardening of the spectrum 
and marginal indication of an increase 
in the iron line strength during the pulse maximum. 
\end{abstract}


\section{Introduction}

\begin{figure}
\psbox[xsize=0.47\textwidth]{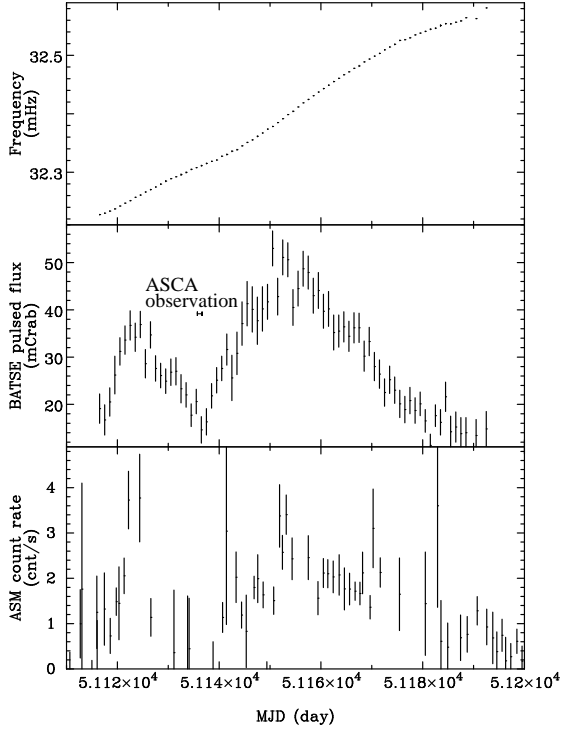}
\caption{History of the pulse frequency and flux. (Top panel) pulse frequency 
determined with {\it CGRO}/BATSE. (Middle panel) BATSE pulsed flux in 20--50 keV. 
Duration of the ASCA observation is indicated by an ``H''-shaped sign. 
(Bottom panel) {\it RXTE}/ASM count rate. 
\label{fig:initial}
}
\end{figure}

The transient X-ray pulsar XTE~J0111.2--7317 was discovered with the
Proportional Counter Array (PCA) of the {\it Rossi X-ray Timing Explorer}
({\it RXTE}) in 1998 November (Chakrabarty et al. 1998a\markcite{Chakrabarty1998a}) 
and was simultaneously detected in hard X-rays 
(Wilson \& Finger 1998\markcite{Wilson1998}) with
the Burst and Transient
Source Experiment (BATSE) onboard the {\it Compton Gamma-Ray Observatory}
({\it CGRO}). 
Public data of {\it CGRO}/BATSE and {\it RXTE}/ASM (All Sky Monitor) 
revealed that 
both the hard and the soft X-ray intensities of the source showed two outbursts 
as shown in Figure \ref{fig:initial}. 
Initial BATSE measurements also found the pulsar to
be spinning-up with a very short time scale of 
$P/\dot{P} \sim 20$~yr, 
which is also a confirmation that the compact object is a neutron star.
The spin-up rate shows a positive correlation with
the flux, which is similar to other transient X-ray pulsars. 

Following its discovery, a {\it Target of Opportunity} observation was made
with the {\it Advanced Satellite for Cosmology and Astrophysics} ({\it ASCA})
and the position of the X-ray source was determined precisely
(Chakrabarty et al. 1998b\markcite{Chakrabarty1998b}).
Previous observations of the field with {\it EINSTEIN} and 
{\it ROSAT} did not detect any source at the position determined
with {\it ASCA} (Wang \& Wu 1992\markcite{Wang1992}; 
Kahabka \& Pietsch 1996\markcite{Kahabka1996};
Cowley et al. 1997\markcite{Cowley1997}; 
Kahabka et al. 1999\markcite{Kahabka1999}; 
Schmidtke et al. 1999\markcite{Schmidtke1999}), 
confirming its true transient nature. 
Since transient pulsars in our Galaxy are located in the Galactic plane 
whereas XTE~J0111.2--7317 is located 
in the direction of the Small Magellanic Cloud (SMC), 
i.e. at high Galactic lattitude ($\sim -43^\circ$), 
it is very likely that XTE~J0111.2--7317 lies in the SMC. 
A candidate for the optical counterpart was detected in the 
{\it ASCA} error circle, and was found to be 
emitting strong H$\alpha$ and H$\beta$ lines 
(Israel et al. 1999\markcite{Israel1999}; 
Coe et al. 1999\markcite{Coe1999}). 
Coe et al. (1999\markcite{Coe1999}) also determined 
the velocity shift of optical emission lines to be 
$166 \pm 15$~km~s$^{-1}$, which indicates that 
this object is in the SMC. 
Therefore the distance would be large (65~kpc is assumed in this paper), 
which indicates a very high luminosity, in excess of
10$^{38}$~erg~s$^{-1}$ in the 2.0--10.0~keV band during the transient phase.

The {\it ASCA} observation of the source was made to study the
pulsations and the soft X-ray spectrum extending upto 10~keV.
Though the {\it ASCA} observation happened to coincide with the quiescent
period between the two outbursts (Figure \ref{fig:initial}), 
the source was bright enough for
us to make a detailed study of its energy spectrum. We present here 
results of 
our temporal and spectral study 
of this source made with the {\it ASCA} data. The principal
result from this study is discovery of a soft excess in the spectrum
which is modeled as an ``inversely broken power-law model.''
In the following sections, we present the details
of the observation and the temporal and spectral analysis, and discuss
the results suggesting that this source is a Be/X-ray binary.

\section{Observation and Analysis}

The observation of the new transient X-ray pulsar XTE~J0111.2--7317 
in the SMC region 
was made between 1998 November 18.73 UT and November 19.65 UT with
two Solid-state Imaging Spectrometers (SIS) and two Gas Imaging
Spectrometers (GIS) of {\it ASCA}. The SIS CCDs were operated in 1-CCD
Faint/Bright mode with time resolution of 4~s, 
and the GIS detectors were operated in normal PH mode
which has higher time resolution of 62.5 ms. For details about 
{\it ASCA} itself,
its X-ray Telescope (XRT), and the focal plane instruments SIS and GIS, 
please refer to Tanaka, Inoue, \& Holt (1994)\markcite{Tanaka1994}, 
Serlemitsos et al. (1995)\markcite{Serlemitsos1995}, 
Burke et al. (1994)\markcite{Burke1994}, and 
Ohashi et al. (1996)\markcite{Ohashi1996}.
The observation resulted in 19.8 ks and
25.7 ks of useful exposures with the SIS and GIS detectors respectively.
Average source count rates in the four {\it ASCA} detectors are 3.5 (SIS0), 2.7
(SIS1), 1.7 (GIS2), and 2.3 (GIS3) per second. The differences in the
count rates between the two identical SIS or GIS detectors are due to
different mirror efficiencies for the nominal positions. 
The source coordinates obtained from the SIS data are, 
R.A. = $01^{\rm h} 11^{\rm m} 14\fs 5$,
Dec = $-73^\circ 16' 50''$ (epoch 2000), 
with an uncertainty radius of $30''$ at 90\% confidence level 
(Chakrabarty et al. 1998b\markcite{Chakrabarty1998b}).

All data were collected when the satellite was outside
the South Atlantic Anomaly region and 
deviation from nominal pointing was less than 0\fdg 01. 
We have used the standard data reduction criteria of the {\it ASCA} guest
observer facility by applying 
minimum cut-off rigidity of 6/4 GeV/$c$ 
and minimum elevation angle of 10$^\circ$/5$^\circ$ 
for SIS/GIS, respectively. 
Minimum angle from the bright earth's limb of 20$^\circ$ was also applied 
for the SIS data. 
Data from the hot and flickering pixels
of the SIS detectors were removed. Charged particle events were removed
from the GIS data 
based on the rise time discrimination method. The count rates and spectra
of the source were obtained from circular regions of $6'$ diameter for
the GIS detectors and $4'$ diameter for the SIS detectors, and the
background data were obtained from sections of the rest of the field of view.

\subsection{Timing analysis}

To study the temporal properties, we have produced light curves from the 
two GIS detectors with 0.5~s time resolution. Arrival times of the
photons were corrected to the solar system barycenter. Data from the two
detectors were added and a power spectrum was generated. Pulsations were
detected unambiguously with a barycentric period of 30.9497 $\pm$ 0.0004~s.
The pulse profile is predominantly sinusoidal with indication of one
additional peak. We also have generated pulse profiles in three different
energy bands of 0.7--2.0, 2.0--5.0, and 5.0--10.0~keV. The pulse
profiles in these energy bands plotted in Figure \ref{fig:pp} 
show slight differences
in shape and larger modulation at higher energy. The pulsed fraction, defined
as the ratio of the pulsed intensity to the total intensity in the entire
energy band of 0.7--10.0~keV, is found to be 25\% and 29\% with the SIS and GIS, 
respectively. 
The pulsed fractions in different smaller energy bands with
the SIS and GIS are 17\% and 24\% in 0.7--2.0~keV range, 27\% and 30\% in
2.0--5.0~keV range, and 28\% and 31\% in 5.0--10.0~keV range, respectively. 
The error of the pulsed fraction is estimated to be $\sim 1$--2\% in each band. 
The difference in the pulsed fraction between SIS and GIS 
would be mainly due to the fact that 
the pulsed fraction is larger at higher energy and the relative detection
efficiency of GIS to SIS increases with energy. 
The poorer time resolution of SIS may also be responsible for 
the difference of pulsed fraction. 
The smaller pulsed fraction at lower energy suggests 
spectral hardening during the pulse peak.

We also searched for variation of a time scale 
from $\sim {\rm minute}$ to $\sim {\rm hour}$ in the light curve, 
but no significant variation was found. 

\placefigure{fig:pp}

\subsection{Spectral analysis}
\subsubsection{Phase-averaged spectra}
After appropriate background subtraction, spectra from the two pairs of
SIS and GIS detectors were combined and the two resultant spectra were
fitted simultaneously with slightly different normalizations for the SIS and GIS.
The energy ranges chosen for spectral fitting are 0.55--10.0~keV for the
SIS and 0.7--10.0~keV for the GIS. 
We first applied several conventional single-component models, 
such as a simple power-law, a broken power-law, and 
a power-law with exponential cutoff, with a neutral absorber. 
We found that these models were inadequate to represent the spectra well: 
reduced $\chi^2$ was $\sim 2.77$ for $\sim 840$ degrees of freedom, 
because the residuals to the model fit showed presence of a soft excess 
below $\sim 2$~keV 
and indication of an emission line around 6.4~keV. 
We also found that the {\it ASCA} bandpass does not extend 
high enough (20--30~keV) to detect the type of spectral break 
observed in most X-ray pulsars.

\begin{figure}
\psbox[xsize=0.47\textwidth]{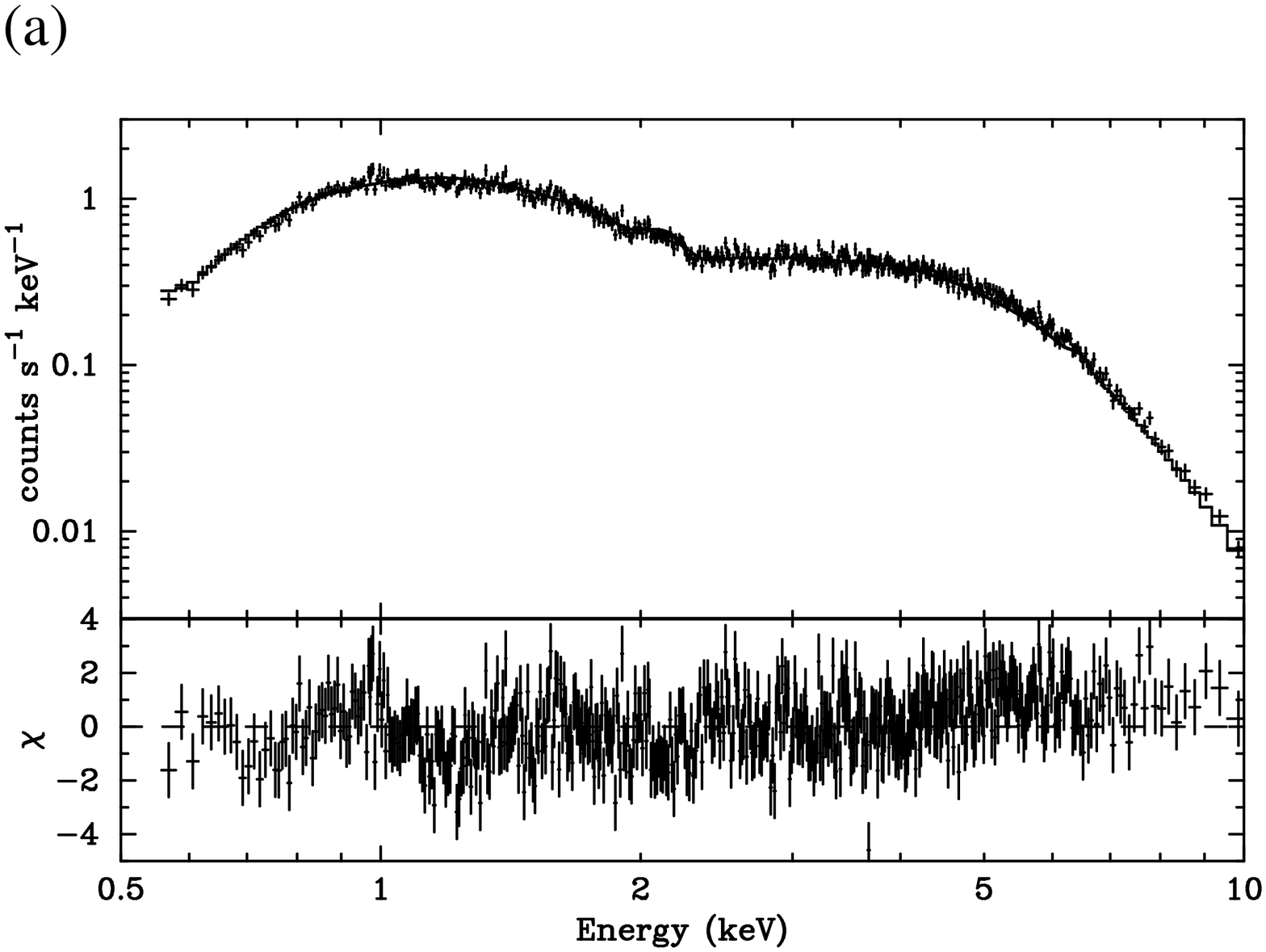}
\psbox[xsize=0.47\textwidth]{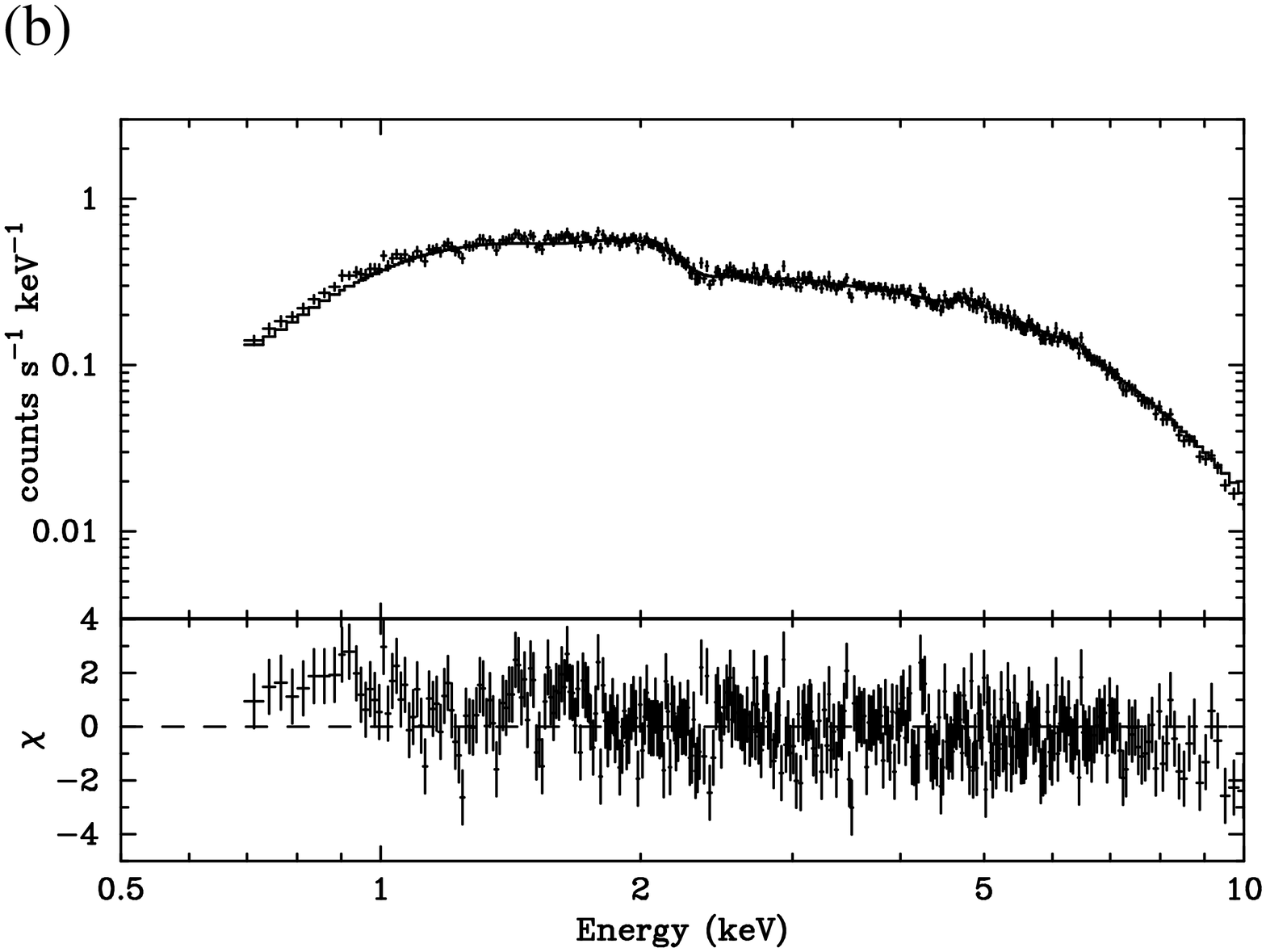}
\caption{Energy spectra of XTE~J0111.2--7317 (crosses) plotted with
the best fitted incident spectra (solid lines) folded through the detectors response
matrices of SIS (a) and GIS (b). 
Spectra from the pairs of SIS and GIS detectors were added to
improve the statistics. The lower panels show the contribution of the
residuals towards the $\chi^2$ at each energy bin.
\label{fig:sp}
}
\end{figure}

\begin{figure}
\psbox[xsize=0.47\textwidth]{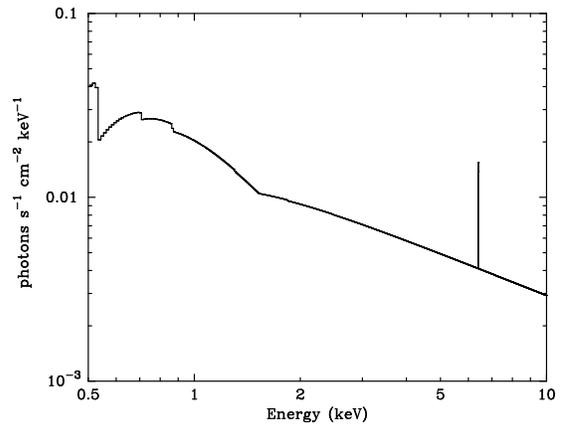}
\caption{Model spectrum with the best-fit parameters. 
\label{fig:model}
}
\end{figure}

The emission-line feature was well described by a narrow gaussian centered at 6.4~keV. 
The continuum shape including the soft excess could be 
represented by an ``inversely broken power-law,'' 
which is a broken power-law with a larger/smaller photon index 
($\Gamma_1/\Gamma_2$)
below/above a break energy $E_{\rm b}$. 
Thus the model of incident photons $f(E)$ follows these equations: 
\begin{eqnarray*}
f(E) &=& e^{-\sigma(E) N_{\rm H}} 
         \left(f_{\rm P}(E) + f_{\rm Fe}(E)\right) \\
     & &   {\rm \ \ \ \ \ [photons\ s^{-1}\ cm^{-2}\ keV^{-1}]}, \\
f_{\rm P}(E) &=& \left\{
                 \begin{array}{@{\,}l@{\,}}
                 I_{\rm P} E^{-\Gamma_1} \ (E < E_{\rm b})   \\
                 I_{\rm P} E_{\rm b}^{(\Gamma_2 - \Gamma_1)}E^{-\Gamma_2} \ 
                 (E \geq E_{\rm b};\ \Gamma_2 < \Gamma_1)   \\   
		 \end{array}
                 \right\},\\
f_{\rm Fe}(E) &=& \frac{I_{\rm Fe}}{\sqrt{2 \pi \sigma_{\rm Fe}^2}}
                  \exp{[-\frac{(E-E_{\rm Fe})^2}{2\sigma_{\rm Fe}^2}]}, 
\end{eqnarray*}
where $E$ and $\sigma(E)$ are the incident photon energy 
and a photoelectric absorption cross-section 
(Morrison \& McCammon 1983\markcite{Morrison1983}), respectively. 
We treated $N_{\rm H}$, $I_{\rm P}$, $\Gamma_1$, $\Gamma_2$, 
$E_{\rm b}$, and $I_{\rm Fe}$ as free parameters, 
while $\sigma_{\rm Fe}$ and $E_{\rm Fe}$ were fixed to be 
zero and 6.4~keV, respectively. 
Best-fit parameters for this model are listed in Table 1. 
Figure \ref{fig:sp} shows the observed spectra together with the best-fit model, 
while the deconvolved model spectrum is shown in Figure \ref{fig:model}. 

\placefigure{fig:sp}
\placefigure{fig:model}
\placetable{tbl-1}

We found that 
two-component models constructed with a hard power-law ($\Gamma \sim 0.77$) 
and a soft component such as a blackbody or a bremsstrahlung, 
which is dominant only below 2~keV, 
could give almost the same fit for the spectra. 
However, these models require a very large emission region 
for the soft component. 
If, in the power-law plus blackbody model, 
the soft component is assumed to be isotropic blackbody emission from a spherical
surface, the corresponding radius is estimated to be $>800$ km, which is
much larger than the neutron star surface area. 
In the power-law plus bremsstrahlung model, on the other hand, 
the emission measure is very high ($>10^{62}$~cm$^{-3}$). 
Taking account of the pulsed fraction below and above 2~keV 
($\sim 20$\% and 30\% respectively), 
and the fact that the soft component in these models is responsible for 
$\sim 70$\% of emission below 2~keV, 
we conclude that the soft component should be pulsating. 
In fact, we extracted on-pulse and off-pulse spectra 
from phase 0--0.5 and 0.5--1, respectively, and 
fitted them with the two-component models mentioned above. 
The fitting result indicated that the pulsed fraction of the soft compoent
in 0.7--2.0~keV band is at least 10--15\%. 
The large emission region of the soft component could not 
explain the pulsation, thus we rejected these two-component models. 

Another two-component model constructed with two power-laws 
and a common absorption also gave a statistically identical fit. 
In this model, however, the soft power-law is very steep 
($\Gamma \sim 6$) and hence the absorption column density 
became large ($\sim 7\times 10^{21}$~H~cm$^{-2}$). 
This requires a deep absorption edge at around 0.6~keV which is
inconsistent with the SIS spectrum, thus we rejected this model.

\subsubsection{Phase-resolved spectra}
\begin{figure}
\psbox[xsize=0.47\textwidth]{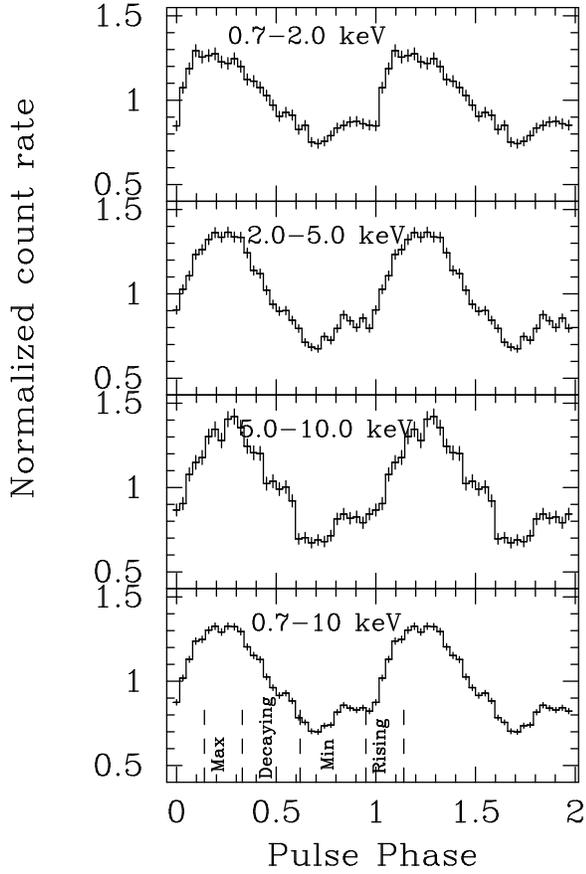}
\caption{Background subtracted pulse profiles obtained with GIS. 
(Top three panels) pulse profiles in small energy bands 
0.7--2.0~keV, 2.0--5.0~keV, and 5.0--10.0~keV. 
A slight change in the shape and amplitude can be noticed. 
(Bottom panel) overall pulse shape in 0.7--10.0~keV 
together with the four pulse regions chosen for the phase 
resolved spectroscopy. 
\label{fig:pp}
}
\end{figure}

The multiband pulse profiles as shown in Figure \ref{fig:pp} indicate 
a slight energy dependence in their amplitude and shape. 
To investigate this in detail, 
we have carried out pulse phase resolved spectroscopy.
We divided the entire pulse phase into four smaller
phases describing the minimum, rising, maximum, and decaying part of the
profile, respectively, as shown in the bottom panel of Figure \ref{fig:pp}.
We fitted the phase-resolved spectra with the same model, and a summary
of the result is given in Table 1.
The photon indices are marginally smaller during the pulse maximum, in accordance
with the energy dependence of the pulsed fraction and the pulse shape in Figure \ref{fig:pp}. 
The iron line equivalent width does not indicate a strong pulse phase dependence.
On the other hand, the iron line normalization
probably indicates an increase during the pulse maximum. 
The break energy and the column density of neutral absorber 
does not indicate a phase dependence.

\section{Discussion}

\subsection{Recent pulsar discoveries}
Recently there has been a significant increase in the discovery of
different types of X-ray pulsars owing to good sky coverage of the
{\it CGRO}/BATSE and {\it RXTE}/ASM for transient sources, 
{\it ASCA} survey of the Galactic ridge, 
and extensive observations with the {\it ROSAT} and {\it BeppoSAX}. 
The first millisecond pulsar SAX J1808.4--3658 showing type-I bursts 
has been the most remarkable discovery 
(Wijnands \& Van Der Klis 1998\markcite{Wijnands1998}; 
Chakrabarty \& Morgan 1998\markcite{Chakrabarty1998c}). 
The other newly discovered pulsars include
a pulsar with type-II bursts, 
pulsars in SGRs (magnetars), pulsars in SNRs, 
anomalous 6-s pulsars and many transient pulsars, 
most of which include a high mass companion. This has improved
the sample of X-ray pulsars by about 40\% in the last 2--3 years
(Nagase 1999\markcite{Nagase1999}). The present source XTE~J0111.2--7317 is one in a
series of new X-ray pulsars discovered in the Magellanic Clouds
(summarized in Yokogawa et al. 2000)\markcite{Yokogawa2000}.

\subsection{Possible origin of the soft excess}

The X-ray continuum spectra of accreting pulsars are often phenomenologically
described to be a broken power-law type or a power-law type with exponential
cutoff (White et al. 1995\markcite{White1995}). 
The break in the spectrum 
for most sources is in the range of 10--20~keV and the power-law
photon index below the break energy is in the range of 1--2.
However, binary X-ray pulsars which are away from the Galactic plane
and therefore experience less interstellar absorption, show the presence of a
soft component in the spectrum which has been often described as a blackbody
and/or thermal bremsstrahlung emission 
(SMC X-1, Woo et al. 1995\markcite{Woo1995}, 
Wojdowski et al. 1998\markcite{Wojdowski1998}; 
LMC X-4, Woo et al. 1996\markcite{Woo1996}; 
RX~J0059.2--7138, Kohno, Yokogawa, \& Koyama 2000\markcite{Kohno2000}; 
4U 1626--67, Orlandini et al. 1998\markcite{Orlandini1998}).
A thermal soft component in 4U 1626--67
requires the size of emission region to be comparable to that of the
neutron star, because the intrinsic luminosity of this source is of the
order of 10$^{35}$ erg s$^{-1}$. But for the pulsars in the Magellanic
Clouds for which the distance is of the order of 50--60 kpc (and the luminosity
is close to the Eddington value), a soft component 
will require an emission region which is a few
orders of magnitude larger than the size of the neutron star. The
bremsstrahlung component in LMC X-4, which is dominant in the intermediate
energy range of 0.5--1.5~keV, is also found to be pulsating 
(Woo et al. 1996\markcite{Woo1996}).
The pulsating nature of the soft component at least in XTE~J0111.2--7317, 
LMC X-4, and SMC X-1 
is difficult to explain if a thermal 
origin is assumed for the low energy part of the spectrum. 

The inversely broken power-law model that we have proposed for
XTE~J0111.2--7317 can possibly explain the pulsations in the
soft X-ray band. If X-rays are emitted from different heights of
an accretion column above the polar cap, then harder X-rays would
be generated in a lower part of the accretion column, because
larger gravitational energy is released there. 
In this situation, larger pulsed fraction could be naturally 
expected in the harder X-ray band, due to the proximity of the emission region 
to the neutron star surface. 
This explanation is qualitatively consistent with 
the observed properties, such as 
the smaller pulsed fraction and the broader peak 
in the lower energy (see Figure \ref{fig:pp}).

{\it ASCA} 
observations have also revealed the presence of a soft excess 
dominated by several emission lines near 1~keV, 
in some high-mass X-ray binary pulsars 
(Vela X-1, Nagase et al. 1994\markcite{Nagase1994}; 
Cen X-3, Ebisawa et al. 1996\markcite{Ebisawa1996}; 
GX 301--2, Saraswat et al. 1995\markcite{Saraswat1995}).
In these sources, pulsation is nearly absent in the low energy band
leading to the conclusion that the soft component is originated at a large
distance from the neutron star, probably due to scattering of the power-law
component by circumstellar material (Nagase et al. 1994\markcite{Nagase1994}). 
The absence of low energy spectral lines in XTE~J0111.2--7317 and the presence of
pulsations at low energy make it different from the pulsars mentioned above.

\subsection{Pulse phase dependence of the spectral shape} 
The fluorescent iron emission line has been detected in many pulsars, and
the equivalent width of the line is found to be upto a few hundred eV in
some cases (White, Swank, \& Holt 1983\markcite{White1983}). 
The iron line detected with an equivalent width of 50 eV in
XTE~J0111.2--7317 is only modest. 

In the pulse phase resolved spectra, we
have found indication of an increase in the iron line strength during the
pulse maximum. Pulse phase resolved spectroscopy of many pulsars indicates
an anti correlation between the iron line equivalent width and the intensity
(White et al. 1983\markcite{White1983}), 
which is consistent with the iron line strength
being constant. {\it GINGA} observations have indicated a pulse phase
dependence of the iron line strength in accretion powered binary pulsars
Cen X-3 (Day et al.  1993\markcite{Day1993}), 
Vela X-1 (Choi et al. 1996\markcite{Choi1996}), and LMC X-4
(Woo et al. 1996\markcite{Woo1996}). {\it GINGA} and
{\it ASCA} observations of GX 1+4 suggest a constant intensity iron line
which shows a large increase in the equivalent width during the pulse
minimum (Dotani et al. 1989\markcite{Dotani1989}; 
Kotani et al. 1999\markcite{Kotani1999}). However, a detailed
study of the pulse phase dependence of an iron line equivalent width in the
X-ray pulsars observed with {\it ASCA} has not been done. 

The slight 
hardening of the spectrum observed in XTE~J0111.2--7317 during the pulse
maximum is common to many accreting X-ray pulsars 
(White et al. 1983\markcite{White1983}).

\subsection{Energy dependence of the pulse shape}
In Figure \ref{fig:pp} we see that the main pulse which arrives at a phase 0.45
earlier than the minimum contributes to most of the observed pulsed
emission. The secondary peak is at a phase 0.17 after the minimum.
The pulse profiles of X-ray pulsars depend on the beaming pattern
(fan or pencil), angles between the magnetic axis and the spin axis
($\theta_m$), and that between the observer line of sight and the spin
axis ($\theta_r$). If $\theta_m + \theta_r > \pi/2$, a multi-peak
profile will be observed because in a pencil beam geometry, emission
from both the polar regions will reach the observer during one revolution
of the neutron star, and in a fan beam geometry, the line of sight will cross
the magnetic equator twice in one revolution 
(Wang \& Welter 1981\markcite{Wang1981}). 
However, for such simple explanation to be valid,
in pencil beam geometry, the two pulse peaks need to be separated by
0.5 in phase and the two minima should have same intensity level, while
in fan beam geometry, the two peaks should have identical strength.
Absence of such symmetric features in the pulse profile indicates that
the emission pattern is complex and also the two magnetic poles may
not be positioned diametrically opposite to each other. A model of pulse
profiles, which includes gravitational light bending and a polar cap or
a ring type emission zone (Leahy \& Li 1995\markcite{Leahy1995}) 
also does not explain the
pulse profiles which are sufficiently complex in structure and show time
variation.

\subsection{Is XTE~J0111.2--7317 a Be/X-ray system in the SMC?}
Although the exact distance to this pulsar has not been measured, 
a possible optical counterpart, 
which has been identified as a Be-type star, 
is likely to be located at the distance of the SMC 
(Coe et al. 1999\markcite{Coe1999}). 
In addition, the absorption column density deduced from the spectral analyses 
($\sim 2 \times 10^{21}$~H~cm$^{-2}$; Table \ref{tbl-1}) is 
comparable to the total HI column density in this direction (Luks 1994), 
which also supports that this object is beyond our Galaxy 
and possibly in the SMC.

The characteristics of this pulsar, like its transient nature, hard X-ray
spectrum, high X-ray luminosity, and the rapid 
spin-up property during the transient phase, 
are analogous to the pulsars in binary 
systems with high mass companion (Bildsten et al. 1997\markcite{Bildsten1997}). 
A pulse period of 31~s is also much 
larger than the period of X-ray pulsars in low mass binaries. 
The present outburst is analogous to the giant outbursts
seen in Be/X-ray pulsars characterized by high luminosity and high spin-up
rates. 
Considering all these properties in the X-ray band 
and the existence of a Be star counterpart, 
we suggest that XTE~J0111.2--7317 is very likely to be 
a Be/X-ray binary system. 

The empirical relation between pulse and orbital periods known to
exist among the pulsars in Be/X-ray binaries (Corbet 1984\markcite{Corbet1984}), 
suggesting an orbital
period of 30--50 days, is in agreement with the observed light
curve during the first 70 days after the transient outburst 
(Figure \ref{fig:initial}).
In the public data of {\it CGRO}/BATSE, the pulse period history shows 
deviation from a linear trend which can be due to a changing mass-accretion
rate or an orbital Doppler shift or a combination of  both. If we simply fit
the pulse period history with a linear plus sinusoidal function, we
obtain a binary period of $\sim 48$~day.
This result is consistent with the predicted value mentioned above, 
hence also supports that this pulsar is a Be/X-ray binary system.

However, two anomalies could be noted. 
One is the photon index of 0.8 above $\sim 2$~keV 
obtained from this {\it ASCA} observation, 
which is lower than the typical 
value of $\sim 1.5$ for the Be/X-ray pulsars (White et al. 1983\markcite{White1983}). 
The other is the presence of a soft excess with strong pulsations, 
which is rarely seen in binary pulsars in our Galaxy. 
However, even if a pulsar in our Galaxy had the soft excess 
like XTE~J0111.2$-$7317, 
it could not be observable in most cases 
due to the strong absorption of the order of $10^{22}$~H~cm$^{-2}$.
The Magellanic Cloud pulsars therefore would be very suitable 
for a systematic study of soft X-rays from binary pulsars.

\section{Conclusions}

We have presented the {\it ASCA} observation of the newly discovered X-ray
pulsar XTE~J0111.2--7317 in its bright transient phase. Pulsations at 30.95~s 
were unambiguously detected, and the pulse shape and the pulsed fraction 
are found to have a slight energy dependence.
The energy spectrum in the 0.7--10.0~keV band has a soft excess above 
a simple power-law type model, and can be described by 
an ``inversely broken power-law'' model, 
which is a broken power-law with a larger/smaller photon index below/above 
a break energy.

\begin{deluxetable}{p{4.0cm}lllll}
\small
\tablenum{1}
\tablecaption{Phase averaged and phase resolved spectral parameters.\label{tbl-1}}
\tablewidth{0pt}
\tablehead{
\colhead{}&\colhead{Average}&\colhead{Rising}&\colhead{Maximum}&\colhead{Decaying}&\colhead{Minimum}}

\startdata
Source counts (s$^{-1}$)              & 10.28                  & 11.33                  & 13.68                  & 10.63                  & 8.46             \nl
$\Gamma_1$                            & $2.3^{+0.1}_{-0.2}$    & $2.3^{+0.4}_{-0.4}$    & $2.0^{+0.3}_{-0.4}$    & $2.2^{+0.3}_{-0.2}$    & $2.2^{+0.3}_{-0.2}$ \nl
$\Gamma_2$                            & $0.76^{+0.01}_{-0.02}$ & $0.81^{+0.04}_{-0.04}$ & $0.71^{+0.03}_{-0.03}$ & $0.72^{+0.02}_{-0.03}$ & $0.78^{+0.03}_{-0.03}$ \nl
$E_{\rm b}$ (keV)                     & $1.53^{+0.02}_{-0.02}$ & $1.48^{+0.06}_{-0.04}$ & $1.50^{+0.05}_{-0.05}$ & $1.55^{+0.04}_{-0.04}$ & $1.57^{+0.05}_{-0.04}$ \nl
$N_{\rm H}$\tablenotemark{a}          & $1.8^{+0.3}_{-0.2}$    & $1.9^{+0.7}_{-0.6}$    & $1.6^{+0.6}_{-0.6}$    & $1.7^{+0.4}_{-0.5}$    & $1.6^{+0.5}_{-0.4}$ \nl
$I_{\rm Fe}$\tablenotemark{b}         & $2.1^{+0.6}_{-0.6}$    & $2.4^{+1.5}_{-1.6}$    & $3.7^{+1.7}_{-1.6}$    & $2.3^{+1.1}_{-1.3}$    & $0.9^{+0.9}_{-0.9}$ \nl
Fe-line eq. width (eV)                & $50^{+14}_{-14}$       & $55^{+34}_{-37}$       & $67^{+31}_{-29}$       & $55^{+26}_{-31}$       & $28^{+28}_{-28}$ \nl
Flux (0.7--10.0~keV)\tablenotemark{c} & 3.6                    & 3.8                    & 5.0                    & 3.8                    & 2.9              \nl
Reduced $\chi^2$ (dof)                & 1.393 (842)            & 1.448 (207)            & 1.277 (249)            & 1.200 (291)            & 1.297 (264)      \nl

\tablenotetext{a}{10$^{21}$ atoms cm$^{-2}$}
\tablenotetext{b}{10$^{-4}$ photons cm$^{-2}$ s$^{-1}$} 
\tablenotetext{c}{10$^{-10}$ erg cm$^{-2}$ s$^{-1}$}
\tablecomments{Errors indicate 90\% confidence limits for a single parameter.}
\enddata
\end{deluxetable}

\end{document}